\documentclass[aps,pra,superscriptaddress,nofootinbib,bibnotes]{revtex4}

\usepackage{graphicx}
\usepackage{amsfonts}
\usepackage{amsthm}
\usepackage{color}
\usepackage{bm}
\usepackage{mathptmx}      

\newtheorem{proposition}{Proposition}
\newtheorem{lemma}{Lemma}

\newcommand{\beq}{\begin{equation}}
\newcommand{\eeq}{\end{equation}}

\newcommand{\R}{{\mathbb{R}}}

\newcommand{\K}{{\mathcal{K}}}

\begin{document}

\title{Three Slit Experiments and the Structure of Quantum Theory}


\author{Cozmin Ududec}
\email[]{cududec@perimeterinstitute.ca}
\affiliation{Department of Physics, University of Waterloo, Waterloo, Canada}

\author{Howard Barnum}
\email[]{hbarnum@perimeterinstitute.ca}
\affiliation{Perimeter Institute for Theoretical Physics, 31 Caroline Street N, Waterloo, Canada}

\author{Joseph Emerson}
\email[]{jemerson@uwaterloo.ca}
\affiliation{Department of Applied Mathematics, University of Waterloo,  Waterloo, Canada}


\begin{abstract}
In spite of the interference manifested in the double-slit experiment,
quantum theory predicts that a measure of interference defined by Sorkin and involving various outcome probabilities from an experiment with {\em three} slits, is identically zero.
We adapt Sorkin's measure into a general operational probabilistic framework for physical theories, and then study its relationship to the structure of quantum theory.
In particular, we characterize the class of probabilistic theories for which the interference measure is zero as ones in which it is possible to fully determine the state of a system via specific sets of `two-slit' experiments.
\end{abstract}

\maketitle

\section{Introduction}

The form of interference that is manifested in the double-slit
experiment is one of the most characteristically quantum phenomena, and is often considered to capture the
essence of quantum mechanics \cite{feyn}.  However, in the vast
literature on quantum interference the focus has largely been on describing,
analyzing, or attempting to explain \emph{two-slit} interference,
with little attention paid to the possibility of new and interesting
phenomena arising when more than two slits are involved.
An exception is the pioneering work of R. Sorkin \cite{sorkin},
who introduced a hierarchy of interference-type phenomena associated
with experiments involving multiple slits.  His hierarchy is described
by a sequence of expressions $I_k$, for $k=2,...,\infty$, where each
$I_k$ is defined in terms of the outcome probabilities of a $k$-slit
experiment.  If $I_k$ is nonzero, then the experiment is said to
exhibit \emph{$k$-th order interference}.  Sorkin discovered the
remarkable fact that quantum theory predicts that there is no
third---nor higher---order interference in nature, i.e., only the
lowest-order expression $I_2$ is non-zero.

More recently, work has begun on an experiment testing for the absence of third-order
interference \cite{sinha}.
However, in the absence of a theoretical framework broader than the quantum formalism, it is not clear precisely why
quantum theory does not exhibit higher than second order interference,
or more generally, what characteristic property of a theory (besides the expression $I_3$ being
zero) three-slit experiments are testing.
In this paper we focus specifically on these questions and characterize the structure of probabilistic theories---satisfying a condition on the allowed transformations---for which $I_3=0$.

We adapt Sorkin's third-order interference expression---originally
defined in a space-time histories and measures language---to a rather
general framework for physical theories in which the primitives of
description are preparation and measurement procedures, and the state
of a system is represented by a vector of probabilities of measurement
outcomes.  Our result characterizes theories which exhibit third-order
interference as ones for which states conditioned on all three slits
being open (i.e., states of systems that have passed the three slits)
cannot be written as linear combinations of states conditioned only on
one or two of the slits being open.  The additional components of
these states can be interpreted as higher-order analogues of the
off-diagonal elements of a density matrix---often called
`coherences'---which are related to interference in two-slit
experiments.  An interesting corollary of this characterization is
that the lack of third-order interference is \emph{equivalent} to the
possibility of doing tomography---asymptotically convergent
statistical estimation of a preparation---via specific sets of `two-slit filtering' experiments.

\section{Quantum Three Slit Experiment}

Consider the idealized setup shown in Fig. $1$, where we have a source
of independently and identically prepared (for simplicity take spin-$1$) systems, with
the spin degrees of freedom of each system described by a possibly
un-normalized state $\rho$. Each system is then sent through a
Feynman filter\footnote{A Feynman filter consists of three
  Stern-Gerlach magnets in series.  The magnets at each end are
  identical, while the middle one is twice as long and has reversed
  polarity.  A beam of spin-$s$ particles is first split into $2s+1$
  spatially separated beams, which are then brought back together into
  one beam upon leaving the apparatus.  A set of internal gates/detectors (one for each
  separated path) can be introduced in the middle magnet. This gives the possibility to filter the beam in
  many different ways by either blocking a path or post-selecting on some detector(s) not firing.}
  \cite{feyn,tomo} aligned along the $\vec{b}$ direction.  After passing the
filter, the systems are measured using a standard Stern-Gerlach magnet
aligned along the $\vec{d}$ axis, together with three detectors
$\{d_l\}_{l=1}^3$.  We represent this measurement with the POVM
$\{\hat{D}_{l}\}_{l=1}^3$, where a positive outcome of the effect
$\hat{D}_{l}$ is associated with the detector $d_l$ firing.

\begin{figure}[h]
    \centering\includegraphics{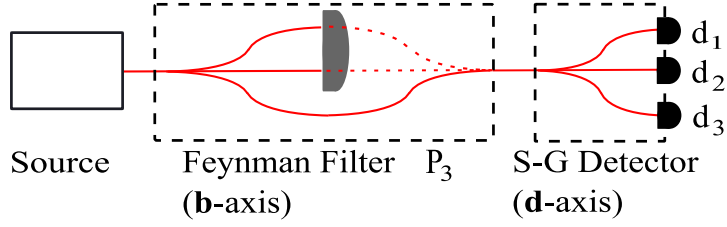}
    \caption{Schematic representation of a `three-slit' experiment based on Stern-Gerlach devices.  A system emitted by the source is sent through an apparatus which acts as a filter (from a specially chosen set).  The systems that pass this filter are then subjected to a standard Stern-Gerlach measurement.}
\end{figure}

Given that we are using spin-$1$ systems, there are seven possible
(nontrivial) Feynman filters: one where we do not filter at all, three
where we block one of the paths, and three more where we block two
paths.  Let $P_{J}$ denote the device constructed by \emph{leaving
  open} the path(s) indexed by $J$, where $J \subseteq \{1,2,3\}$ and
$J \neq \emptyset$, and let the projection operator $\hat{P}_{J}$
represent the transformation implemented by this filter.
Further, let $e_{J}$ represent the experimental event ``the system
passed the filter $P_{J}$.''

The probability that a system will pass
the filter $P_{J}$ is given by $prob(e_{J} | \rho) = Tr[
  \hat{P}_{J} \rho]$.  Further, the joint probability that a system
passes the $P_{J}$ filter \emph{and then} the detector $d_l$ fires is
given by \beq\label{joint1} prob(d_l \ \& \ e_{J} | \rho) = prob(d_l |
e_{J} \ \& \ \rho) prob(e_{J} | \rho)= Tr[\hat{D}_{l}\hat{P}_{J}\rho
  \hat{P}_{J}].  \eeq

Given a preparation $\rho$, a set of filters $\{ P_J \}_{J \subseteq \{1,2,3 \}}$, and a detector $d_l$, we define the \emph{third-order interference expression} (with respect to these devices) as:
\beq\label{I3}
I_3[d_l,\{P_{J} \}, \rho]:= prob(d_l \ \& \ e_{123} | \rho) -  \sum_{1 = j<k}^3 prob(d_l \ \& \ e_{jk}| \rho)
+ \sum_{i=1}^3 prob(d_l \ \& \ e_{i}|  \rho)
\eeq

This kind of expression was introduced in \cite{sorkin} by Sorkin, who considered a set of three-slit experiments with electrons,
and superimposed the seven interference patterns by using a plus sign
when an odd number of the slits are open and a minus sign when an even
number are open.  Quantum theory predicts that for all preparations $\rho$, all final effects $\hat{D}_l$ (representing
events $d_l$), and all intermediate sets of projectors $\{ \hat{P}_J
\}_{J \subseteq \{1,2,3 \}}$ which satisfy the relations $\hat{P}_J \hat{P}_K =
\hat{P}_K \hat{P}_J = \hat{P}_{J \cap K}$ (and represent filtering
devices $P_J$), the expression $I_3[d_l,\{P_{J} \}, \rho]$ is
identically \emph{zero}.

An important point to notice about the expression $I_3$ is that each of the terms that make it up is a probability conditioned on a \emph{distinct} set of open and blocked slits.
There is no a-priori reason for this expression (or any other from Sorkin's hierarchy) to be zero.  In the absence of physical input, probability theory does not constrain probabilities conditioned on different experimental situations \cite{ballentine,jaynes}.
Nevertheless, it should not be surprising that a physical theory will in fact constrain probabilities pertaining to related experimental contexts, and quantum theory predicts a very particular relationship between them.  In order to understand the structure of theories satisfying this relationship, we first need to abstract the essential elements of the above considerations into a setting more general than the quantum formalism.

\section{Probabilistic Models}

We briefly review the necessary parts of the operational probabilistic
framework for physical theories
\cite{mackey,holevo,gudder,hardy,mana2}\footnote{For
  quantum and classical examples of the following concepts and
  mathematical objects, see in particular \cite{hardy} and
  \cite{mana2}. }.  In this framework, the primitive elements are
experimental devices and statistics.  Some devices are taken to act as
\emph{preparations} of a system, and others as \emph{operations} on
the system.  With each use, an operation device performs one of a
possible set of \emph{transformations} on a system, where each
transformation can occur with some probability.  The occurrence of
each transformation is identified by a distinct macroscopic
\emph{event} or \emph{outcome}.  For each pair of a preparation device
$S$ and an operation device $O$, the probabilities $prob(i | O,S)$ for
each outcome $i$ associated with $O$ are assumed be predicted by the theory.

More precisely, a---type of---physical system is modeled by a pair
$(\K, u)$ of a \emph{positive cone} $\K \subseteq V$, which
is a set closed under positive scalar multiplication and addition, and
which spans the \emph{real} vector space $V$ but contains no nontrivial
subspace of $V$.
Each preparation device $S$ is represented by an
element $s \in \K$.  In this manner $\K$ is regarded as consisting of
the un-normalized \emph{states} of a system.  We will restrict
attention to the case where each preparation requires only a
\emph{finite} number of real parameters to specify, so $V=\R^m$ for
some $m < \infty$.
The \emph{order unit} $u$ is a linear functional
on $V$ which is strictly positive on non-zero elements of $\K$, and
defines the set of \emph{normalized} states $\Omega := \{s \in \K |
\ u (s) =1 \}$.

Events/outcomes are represented by linear functionals $e \in V^*$
which satisfy $0 \leq e (s) \leq 1$ for all $s \in
\Omega$, and are often called \emph{effects}.  The set of all effects
on the cone $\K$ will be denoted by $[0,u]$.  The probability that an
event represented by $e$ will occur when the system is prepared by a
device represented by $s$ is then given by $prob(e|s)=e(s)$.

In finite dimension, one can use an isomorphism between $V$ and $V^*$ to embed the set of effects $[0,u] \subset V^*$ into the space $V$ containing the states.  This can be used to define an inner product between effects and states, which we will use to represent the above probabilities as $prob(e|s)=e \cdot s$.
This inner product can be interpreted as a `Born rule', which is thus seen to be valid and fundamental for all theories in this framework.
The quantum Born rule, $prob(e|\rho)=Tr(\hat{E} \rho)$, is a particular representation \cite{hardy} of the above inner product that results from the particular (quantum) geometry of states and effects.
It is the \emph{geometry} of the spaces of states and effects that defines a theory, and what is being generalized here.

Transformations of a system are represented by linear maps $\phi: V
\rightarrow V$ which satisfy $\phi(\K) \subseteq \K$ and $u \cdot \phi(s) \leq u\cdot s $ for all $s \in \K$, i.e.,
they are \emph{positive} in the sense that they take allowed states to
allowed states, and are also normalization non-increasing.  An
operation is then represented by a family $\{ \phi_i\}_{i=1}^n$ of
transformations which satisfy $\sum_i^n u \cdot \phi_i(s) =u \cdot s $ for all $s\in \K$.  Since the occurrence of a
transformation is always identified by some event, we define the
effect $e_{i}$ associated with the transformation $\phi_i$ by the
action $ e_{i} \cdot s:=  u \cdot \phi_i(s)$, for all $s\in \K$.
We regard $ e_{i}\cdot s / u \cdot s $ as the probability that the $i$-th outcome occurs when the state is $s$.
The (normalized) state of a system conditioned on the outcome $i$ having occurred is then given by $s_i = \frac{u \cdot s}{e_i \cdot s} \phi_i(s)$.

A set of effects $\{e_i\}_{i=1}^n$ which satisfy $\sum_{i=1}^n e_i =u$ is called a \emph{measurement}.  If the probabilities associated with all the outcomes in a measurement are sufficient to uniquely determine any given state, then it is called an \emph{informationally complete measurement}.  All finite dimensional models support an informationally complete measurement \cite{barrett}.

The concepts of exposed faces and filters have played an important role in many axiomatizations of quantum theory \cite{AandS,araki,ludwig,mielnik} and will be essential in what follows.
A convex subset $F$ of a cone $\K$ is called a \emph{face} if it is closed under convex combination and decomposition.  An \emph{exposed face} is a face which has the further property that there exists some effect $f$ such that $F = kernel(f) \cap \K$.
As an example, the faces and the exposed faces of a quantum mechanical model coincide and correspond to the subspaces of the Hilbert space.

A \emph{filter} $P$ is a transformation with the following properties:
\begin{enumerate}
\vspace{-\topsep}
\renewcommand{\theenumi}{\roman{enumi}}
\renewcommand{\labelenumi}{(\textit{\theenumi})}
\item{$P$ is a projection: $P P = P$,}
\vspace{-\topsep}
\item{$P$ is neutral: $ u\cdot P(s)  = u\cdot s $ implies $P(s)=s$,}
\vspace{-\topsep}
\item{$P$ is complemented: there exists at least one neutral projection $P'$ such that $P(s)=s$ if and only if $P'(s)=0$, and $P'(s)=s$ if and only if $P(s)=0$, for all $s \in \K$.}
\vspace{-\topsep}
\end{enumerate}
The interpretation of a filter is that of an idealized type of
transformation, where the first property above represents the
requirement that the state of a system which has been acted on by a filter will be
unchanged if it passes through that type of filter again.  The second
property states that filters are `minimally disturbing' in the sense
that they do not affect systems which they transmit with probability
one.  The third property represents the requirement that for every
filter there is another filter which acts as a `negation' in the sense
that the set of states that pass the filter $P$($P'$) unchanged is
identical to the set of states which do not pass the filter $P'$($P$).

For the remainder of the paper we will focus on a class of models
$(\K,u)$ which satisfy the following:\\
\textbf{Standing Condition:} \emph{Each exposed face $F$ of $\K$ has a unique filter $P_F$ associated with it such that $F=\{s \in \K | \ P_F(s)=s\}$.
Further, the complement of $P_F$ is also unique.}

This condition expresses the requirement (which is satisfied in both classical and quantum theory) that for each exposed face of a model, there is only one filter which transmits those and only those states without affecting them, and only one filter which does not transmit those and only those states.
Given this condition,
 the set of all exposed faces and the
set of all filters of a model are (isomorphic) orthomodular lattices
(see \cite{AandS,belt} for definitions and proofs).  The
lattice operations on pairs of faces $F, G$ are given by $F\wedge G:= F\cap G$ (the largest faces \emph{contained in} both $F$ and $G$) and $F \vee G := (F' \cap G')'$ (the smallest face \emph{containing} both $F$ and $G$), where $F':=\{s \in \K | \ P_F(s)=0\}$.  These operations then induce lattice operations on the
set of filters in the obvious manner\footnote{The standing condition is equivalent (in finite dimensions) to the assumption of
  `spectral duality' used in \cite{AandS} as part of a
  characterization of Jordan-Banach algebra state spaces.  Further,
  orthomodular lattices have played a large role in the quantum logic
  tradition and are closely related to requirements on `conditional
  probabilities' \cite{AandS,belt}.}
.

\section{The Structure of Models With No Third-Order Interference}

The transition from the quantum three-slit experiment discussed in Section $2$ to a generalized `three-slit' experiment is now simple: take the quantum devices and mathematical objects representing them, and replace these with preparations and operations from any other model $(\K,u)$ which satisfies the standing condition.  More precisely, instead of an initial quantum state $\rho$, we take a state $s \in \K$, and instead of quantum effects $\hat{D}_l$, we take effects $r_l \in [0,u]$.
For the generalization of the Feynman filters we take black box devices denoted by $P_{J}$, which simply have the
properties of filters on the state space $(\K,u)$.
The probability that a system passes the  filter $P_{J}$ and then the detector $r_l$ fires is now given by $prob(r_l \ \& \ e_J | s) = r_l \cdot P_J (s)$.

A final and essential prerequisite for formulating a non-trivial three-slit experiment is that the model $(\K,u)$ support a triple of filters $\{P_1, P_2, P_3\}$ which satisfy $P_i P_j =P_j P_i=\delta_{ij}P_i$ for all $i,j=1,2,3$.
Using such a triple (which we take to represent the three `single-slit' experiments) we generalize the multiple-slit filters from the quantum experiment by taking $P_{ij}:=P_i \vee P_j$, and $P_{123}:= P_1 \vee P_2 \vee P_3$.  The pairwise orthogonality of the $P_i$ together with the standing condition will ensure that these $P_{J}$ are in fact filters which satisfy
\beq\label{filters}
P_{J}P_{K}=P_{K}P_{J}=P_{J \cap K}.
\eeq

The above requirement on the transformations representing the single slits generalizes the idea that systems that pass a particular single-slit filter should be perfectly distinguishable from systems that pass another single-slit filter.
It can also be seen as a translation of Sorkin's requirement that the sets of histories that pass through distinct single-slits should be mutually disjoint.
Further, the definition of the $P_J$ and the implied multiplicative properties expressed by (\ref{filters}) capture what is essential in the usual notion of an idealized multiple-slit experiment, and in particular, the operational meaning of leaving two or more slits open in the experiment.

The third-order interference expression is now given by
\beq\label{3rdorder}
I_3[r_l,\{P_J\},s]= r_l \cdot [P_{123} - P^{(3)}](s),
\eeq
where  $P^{(3)} := P_{12} + P_{13} + P_{23} - P_1 - P_2 - P_3$.
The following proposition characterizes models with no third-order interference in terms of the operators $P_{123}$ and $P^{(3)}$, and the relationship between the faces---of filtered states---defined by $F_{J}:=\{ s \in \K | \ P_{J}(s)=s \}$.
\begin{proposition}\label{I3result}
Let the state space $(\K,u)$ satisfy the standing condition.  Take a triple of pairwise orthogonal filters $\{P_{i}\}_{i=1}^3$, and the set of filters $\{P_J \}_{J \subset \{1,2,3\} }$ generated by this triple.  Then the following are equivalent:
\begin{enumerate}
\vspace{-\topsep}
\renewcommand{\theenumi}{\roman{enumi}}
\renewcommand{\labelenumi}{(\textit{\theenumi})}
\item{$I_3[q,\{P_J\},s]=0$ for all $s \in \K$ and $q \in [0,u]$,}
\vspace{-\topsep}
\item{$P_{123} = P^{(3)}$,}
\vspace{-\topsep}
\item{$F_{123} \subset lin[F_{12} \bigcup F_{23} \bigcup F_{13}]$.}
\end{enumerate}
\end{proposition}

\begin{proof}
See Appendix.
\end{proof}

The condition that $F_{123} \subset lin[F_{12} \bigcup F_{23} \bigcup F_{12}]$ expresses the property that states conditioned on all three slits being open (i.e., states of systems that have passed the three slits and are therefore in the face $F_{123}$) can be written as linear combinations of states which are conditioned only on two of the slits being open.  This may seem like a mysterious property at first sight, but it has an intuitive and operational interpretation.

\section{Third-Order Interference and Tomography}

Suppose we are given a device that outputs a set of identically and
independently prepared systems, each described by some model
$(\K,u)$ which satisfies the requirements of Proposition $1$.
For simplicity assume that the filter $P_{123}$ acts as the identity on the whole state space under consideration, i.e., $\K=F_{123}$.
Our task is to reconstruct the state $s$ which represents this device by measurements on the systems it outputs.

To accomplish this task we are only allowed to use the following: $(1)$ the three `double-slit' filters $P_{ij}$, and $(2)$ for each $P_{ij}$, a measurement device $M_{ij}$ which is informationally complete for the systems which pass $P_{ij}$.
What we can do (for each of the given filters) is take a sub-ensemble of the systems produced by the source, pass them through $P_{ij}$, and then use the device $M_{ij}$ on the resulting filtered ensemble to determine the state $s_{ij}=P_{ij}(s) \in F_{ij}$.

If the model we are studying is quantum mechanical, then the
information gained from this filtering and measuring procedure
will in fact be sufficient to find a density matrix which describes the source \cite{tomo}.  In other words, in order to specify a $d \times d$ density matrix (where $d=rank(P_{123})$), it is sufficient to do tomography on the three subspaces $F_{ij}$ of  filtered states\footnote{In particular, if we assume for simplicity that the three filters $P_i$ are all rank-$1$ projections, then $F_{123}$ is the cone of un-normalized
states of a three-level system ($3 \times 3$ positive
semi-definite Hermitian matrices), and the faces $F_{ij}$ correspond
to the ranges of rank-$2$ projections on $F_{123}$ ($2 \times 2$ positive
semi-definite Hermitian matrices).  So in order to specify a $3 \times 3$ density matrix, it is sufficient to do tomography on three two-dimensional subspaces of filtered states.}.

The sufficiency of this kind of tomography for quantum theory
generalizes to all models $(F_{123},u)$ satisfying
$I_3[q,\{P_{J}\},s]=0$ for all $s \in F_{123}$ and $q \in [0,u]$.  In
other words, if the model $(F_{123},u)$ exhibits no third-order
interference (with respect to the experiments defined by the filters $\{P_{J}\}$),
then the components of a state $s \in F_{123}$ can be reconstructed from the
measurements $M_{ij}$ on the faces $F_{ij}$ of filtered states.  This
follows from condition (iii) of Proposition $1$, together with the
fact that the $P_{ij}$ do not disturb the states which they transmit with probability one.
The reconstruction formula for the state $s$ in terms of the filtered states $s_{ij} \in F_{ij}$ is given by
\beq
s = s_{12} + s_{13} + s_{23} - s_{1} - s_{2} - s_{3},
\eeq
where the $s_i:= P_i(s_{ij}) \in F_{ij}$ can easily be inferred once the $s_{ij}$ are determined experimentally.

Conversely, if for some model $(F_{123},u)$ which satisfies the conditions of Proposition $1$, the components of a state $s \in F_{123}$ can be reconstructed from some measurements $M_{ij}$ on states filtered by the $P_{ij}$, then this model will not exhibit third-order interference (with respect to the experiments defined by the filters $\{P_J \}$).
Further, if the state space $F_{123}$ consists of filtered states from a larger state space $\K$, i.e., $F_{123}=P(\K)$ for some filter $P$, then the model $(\K,u_{\K})$ will not exhibit third-order interference either.

On the other hand, if a model \emph{does} exhibit third-order interference, then there are extra parameters which are needed to describe states filtered by $P_{123}$ over and above all the parameters needed to describe states filtered by each of the $P_{ij}$.
Operationally this means that there are measurements which can be performed on states $s \in F_{123}$, the outcome probabilities of which cannot be determined from knowledge only of the outcome probabilities of all possible measurements on the filtered states $s_{ij}$.
The additional parameters can be interpreted as higher-order analogues of the off-diagonal elements of a density matrix---often called `coherences'---which are responsible for interference in two-slit experiments.

\section{Discussion}

We have studied Sorkin's third-order interference expression in the
setting of operational probabilistic models.  We showed that, given a
condition on the kinds of filters a model supports, the absence of
third-order interference is equivalent to the possibility of
reconstructing a state via specific sets of `two-slit filtering'
experiments.

This result gives new insight into the structure of quantum theory
and the implications of three-slit experiments.  The presence of third-order interference in a set of experiments implies that more parameters are needed to describe a system than those specified by the quantum formalism\footnote{Of course it might also be taken to suggest that the assumptions of Proposition $1$ fail, or that the devices used in the experiments do not act as proper filters.  These possibilities can be laid to rest by experimental tests however.}.
Our result also suggests a novel way of testing the structural
property---$(iii)$ from Proposition $1$---which determines whether a model exhibits third-order interference: test whether the tomography procedure outlined above is in fact sufficient to fully characterize actual preparations.

One issue we have not discussed\footnote{Many of the following issues and questions will be explored in a forthcoming paper \cite{coz}.}
is the use of filters (with specific
relations holding between them) to represent generalized slits,
as well the role of the standing condition in our result.  It may in fact be
possible to drop the uniqueness requirement, or even to generalize to a broader class of transformations representing the slits.
It would be interesting to further explore what kinds of objects or concepts are needed, or what conditions a theory must satisfy in order to be able to formulate interference experiments.

As for other models which do not exhibit third-order interference, it can be proven that the state spaces of finite dimensional Jordan-Banach
algebras have this property.  These models include real, complex, and
quaternionic quantum mechanics, and have often been the object of
axiomatic characterization \cite{AandS,vNandJandW} as a stepping stone to complex quantum mechanics.

Finally, we have only discussed the $I_3$ level of Sorkin's hierarchy,
but it is possible to extend the form of analysis we have
used to all the other levels.  We can then ask whether for each level
$k$ of the hierarchy there exists a specific kind of filtering
tomography which is sufficient to fully determine states of models satisfying $I_k=0$.  More
generally, it would be interesting to begin a study of how each
interference expression is related to other nonclassical phenomena
that generalized models exhibit, such as information processing
properties, non-locality, symmetry properties, etc..

\begin{acknowledgements}
We would like to thank Lucien Hardy, Luca Mana, and David Ostapchuk for helpful discussions and comments.
This work was supported in part the Government of Canada through NSERC and Cifar, and by the province of Ontario through OGS.  Research at Perimeter Institute is supported by Industry Canada and the Ministry of Research and Innovation.
\end{acknowledgements}

\section{Appendix: Proof of Proposition 1}

First, note that the conditions on $P$ and $P'$ in property
$(iii)$ of the definition of filters can be re-written as $im^+ P=ker^+
P'$ and $im^+ P'=ker^+ P$, where $im^+ P:=image (P) \cap K$ and
$ker^+ P:=kernel (P) \cap \K$.  Further, let $R_{123}:= P_{123}- P^{(3)}$.
The following lemmas will be useful
for the main proof.

\begin{lemma}\label{linim3}
$P^{(3)}$ is a (not necessarily positive) projection.
\end{lemma}
\begin{proof}
Checking that $P^{(3)}P^{(3)}= P^{(3)}$ is a simple exercise in applying the definition of $P^{(3)}$ and then using the fact that $P_KP_J=P_JP_K=P_{J\cap K}$.
\end{proof}

\begin{lemma}\label{linim3}
$lin[F_{12} \bigcup F_{23} \bigcup F_{13}] = im (P^{(3)})$.
\end{lemma}
\begin{proof}
That $im(P^{(3)}) \subseteq lin[F_{12} \bigcup F_{23} \bigcup F_{12}]$, is immediate from the definition of $P^{(3)}$.
We also have that $P^{(3)}P_i = P_i P^{(3)} = P_i$ and $P^{(3)}P_{ij}=P_{ij}
P^{(3)}= P_{ij}$ for all $1\leq i<j \leq 3$, so $P^{(3)}$ acts as the
identity on the subspace $lin[F_{12} \bigcup F_{23} \bigcup F_{12}]$, and therefore
$lin[F_{12} \bigcup F_{23} \bigcup F_{12}] \subseteq im(P^{(3)})$.
\end{proof}

\noindent
\emph{Proof of Proposition 1.}
$(i) \Leftrightarrow (ii)$ is clear from the definitions.
$(ii) \Leftrightarrow (iii)$ It is not difficult to see that $R_{123} P^{(3)}=P^{(3)} R_{123}=0$, and using the fact that $P^{(3)}$ is a projection, we have $im (P_{123}) = im (R_{123}) \oplus im (P^{(3)})$ and $ker (P_{123}) = ker(R_{123}) \cap ker(P^{(3)})$.
These equalities (along with $P^{(3)} R_{123}=0$) imply that $P_{123} = P^{(3)}$ if and only if $R_{123} = 0$.
Combining this with Lemma 1 gives $im (P_{123}) = lin[F_{12} \bigcup F_{23} \bigcup F_{12}]$ if and only if $R_{123}=0$.  Using the facts that $im (P_{123})= im^+ P_{123} - im^+ P_{123}$, and $F_{123}= im^+ P_{123}$, we see that $im (P_{123}) = lin[F_{12} \bigcup F_{23} \bigcup F_{12}]$ is equivalent to $F_{123} \subset lin[F_{12} \bigcup F_{23} \bigcup F_{12}]$.
\qed

\end{document}